\documentclass[twocolumn,showpacs,preprintnumbers,amsmath,amssymb,floatfix]{revtex4}

\usepackage{graphicx}
\usepackage{dcolumn}
\usepackage{bm}

\begin{document}

%

\let\a=\alpha      \let\b=\beta       \let\c=\chi        \let\d=\delta
\let\e=\varepsilon \let\f=\varphi     \let\g=\gamma      \let\h=\eta
\let\k=\kappa      \let\l=\lambda     \let\m=\mu
\let\o=\omega      \let\r=\varrho     \let\s=\sigma
\let\t=\tau        \let\th=\vartheta  \let\y=\upsilon    \let\x=\xi
\let\z=\zeta       \let\io=\iota      \let\vp=\varpi     \let\ro=\rho
\let\ph=\phi       \let\ep=\epsilon   \let\te=\theta
\let\n=\nu
\let\D=\Delta   \let\F=\Phi    \let\G=\Gamma  \let\L=\Lambda
\let\O=\Omega   \let\P=\Pi     \let\Ps=\Psi   \let\Si=\Sigma
\let\Th=\Theta  \let\X=\Xi     \let\Y=\Upsilon

%

%

\def\cA{{\cal A}}                \def\cB{{\cal B}}
\def\cC{{\cal C}}                \def\cD{{\cal D}}
\def\cE{{\cal E}}                \def\cF{{\cal F}}
\def\cG{{\cal G}}                \def\cH{{\cal H}}
\def\cI{{\cal I}}                \def\cJ{{\cal J}}
\def\cK{{\cal K}}                \def\cL{{\cal L}}
\def\cM{{\cal M}}                \def\cN{{\cal N}}
\def\cO{{\cal O}}                \def\cP{{\cal P}}
\def\cQ{{\cal Q}}                \def\cR{{\cal R}}
\def\cS{{\cal S}}                \def\cT{{\cal T}}
\def\cU{{\cal U}}                \def\cV{{\cal V}}
\def\cW{{\cal W}}                \def\cX{{\cal X}}
\def\cY{{\cal Y}}                \def\cZ{{\cal Z}}
%

\newcommand{\Ns}{N\hspace{-4.7mm}\not\hspace{2.7mm}}
\newcommand{\qs}{q\hspace{-3.7mm}\not\hspace{3.4mm}}
\newcommand{\ps}{p\hspace{-3.3mm}\not\hspace{1.2mm}}
\newcommand{\ks}{k\hspace{-3.3mm}\not\hspace{1.2mm}}
\newcommand{\des}{\partial\hspace{-4.mm}\not\hspace{2.5mm}}
\newcommand{\desco}{D\hspace{-4mm}\not\hspace{2mm}}



\title{ Direct CP Violation in $B^+\to J/\psi K^+$ Decay
        as Probe for New Physics
}
\author{Wei-Shu Hou$^{a}$}
\author{Makiko Nagashima$^a$}
\author{Andrea Soddu$^{b}$}
\affiliation{ $^a$Department of Physics, National Taiwan
 University, Taipei, Taiwan 10617, R.O.C. \\
$^b$Department of Particle Physics, Weizmann Institute
 of Science, Rehovot 76100, Israel
}
\date{\today}

\begin{abstract}
Currently there are New Physics hints in mixing-induced $CP$
violation ${\cal S}_{\phi K^0}$, ${\cal S}_{\pi^0 K^0} < {\cal
S}_{J/\psi K^0}$, unequal direct $CP$ violation ${\cal
A}_{K^+\pi^-} \neq {\cal A}_{K^+\pi^0}$, and maybe even in
measured ${\cal S}_{J/\psi K^0}$ vs prediction from global fit to
other data. However, these hints either suffer from experimental
uncertainties, or uncertain hadronic corrections, and are not yet
unequivocal. Motivated by these hints, however, we point out that
a unique probe may be {\it direct} $CP$ violation in $B\to J/\psi
K$ mode. An asymmetry observed at 1\% or higher would indicate New
Physics.
\end{abstract}

 \maketitle


Unprecedented luminosities have been achieved at the asymmetric
energy $e^+e^-$ collider B factories, where altogether close to
$10^9$ $B\bar B$ meson pairs have been collected so far. Besides
establishing $CP$ violation (CPV) in the $B$ system, a few hints
for possible New Physics (NP) have emerged:
the difference between time dependent CPV (TCPV) in charmless
$b\to s\bar qq$ modes vs $b\to c\bar cs$ modes; the difference in
direct CPV (DCPV) asymmetries between $B^0\to K^+\pi^-$ vs $B^+\to
K^+\pi^0$; and maybe even in the slightly lower value for TCPV in
$B^0\to J/\psi K^0$ mode compared with the predicted value from
fits to other data.

All three hints are not unequivocal, and suffer either in
experimental significance, or in theoretical interpretation.
In this Letter we point out that DCPV in the $B^+\to J/\psi K^+$
mode, ${\cal A}_{J/\psi K^+}$, could be at the 1\% level or
higher, {\it if} the above hints are true harbingers of NP.
Current measurements give~\cite{expt} ${\cal A}_{J/\psi K^+} =
0.018 \pm 0.043 \pm 0.004$ (CLEO), $-0.026 \pm 0.022 \pm 0.017$
(Belle) and $0.03 \pm 0.015 \pm 0.006$ (BaBar), based on 9.7M,
31.9M and 89M $B\bar B$ pairs, respectively. Note that this has
not been updated since 2003. The average is~\cite{PDG05}
\begin{eqnarray}
{\cal A}_{J/\psi K^+} = 0.016 \pm 0.016,
 \ \ \ \ \ \ \ \ \ \ \ \ ({\rm PDG}\ 2005)
 \label{eq:expt}
\end{eqnarray}
where a scale factor has been applied to the error. The nominal
error should be of order 0.012 for 131M events, hence a
statistical error of 0.003 should be attainable with $\sim$ 2000M
events expected by 2008.

To improve statistics, one could combine with $B^0\to J/\psi K^0$
mode. Since Standard Model (SM) expectation is at the sub-percent
level, in the next few years ${\cal A}_{J/\psi K}$ could be a
better probe of NP if seen above 1\% level. Such ``precision
measurement" studies would be good preparation for the future
Super B factory, where systematic issues would become a main
concern.


%
Since a few years, a contrast has emerged in mixing-induced TCPV
in $B$ decays, i.e. ${\cal S}_f$ measurement in a host of $CP$
eigenstates $f$.
The current world average of ${\cal S}_f$ in $b\to c\bar cs$
decays is $\sin2\phi_1/\beta = 0.685 \pm 0.032$~\cite{HFAG}, which
is dominated by the $B^0\to J/\psi K_S$ mode.
TCPV measurements in loop dominated $b\to s\bar qq$ processes such
as $B^0 \to \eta^\prime K^0$, $\phi K^0$ and $\pi^0 K^0$, on the
other hand, have persistently given values below
$\sin2\phi_1/\beta$. The current average in $b\to s\bar qq$ decays
is~\cite{HFAG} ${\cal S}_{s\bar qq} = 0.50 \pm 0.06$. The
difference $\Delta {\cal S} = {\cal S}_{s\bar qq} -
\sin2\phi_1/\beta$ is only at 2.7$\sigma$ level. However, $\Delta
{\cal S}$ has been diminishing experimentally in the past two
years, while theoretical interpretation suffers from hadronic
uncertainties~\cite{HNRS}.

DCPV was recently observed \cite{AKpiAKpi0} in $B^0\to K^+\pi^-$
decay, giving ${\cal A}_{K^+\pi^-} = -0.108 \pm
0.017$~\cite{HFAG}. However, no indication was seen in charged
$B^+\to K^+\pi^0$, giving ${\cal A}_{K^+\pi^0} = 0.04 \pm 0.04$.
The difference with ${\cal A}_{K^+\pi^-}$ could be due to an
enhancement of color-suppressed amplitude $C$~\cite{LargeC}, or
due to electroweak penguin $P_{\rm EW}$ effects \cite{Kpi0HNS}.
For the latter, NP CPV phase would be needed. Although the
experimental significance for ${\cal A}_{K^+\pi^0} - {\cal
A}_{K^+\pi^-}$ is relatively robust, at present it is not yet
conclusive whether NP phase is absolutely called for.

A third possible hint, though not yet widely perceived, is the
${\cal S}_f$ value in $b\to c\bar cs$ decays itself. The Belle
experiment gives~\cite{SpsiKBelle} ${\cal S}_{J/\psi K^0} =  0.652
\pm 0.039 \pm 0.020$ using 386M $B\bar B$ pairs, which dominates
the HFAG average of ${\cal S}_{c\bar cs} =  0.685 \pm
0.032$~\cite{HFAG} over charmonium modes. The latter is pulled up
slightly by the BaBar result of ${\cal S}_{c\bar cs} = 0.722 \pm
0.040 \pm 0.023$ based on 227M $B\bar B$ pairs. Early Belle
results based on a smaller data sample also averaged over many
charmonium modes, and gave a higher value. It remains to be seen
what would be the future update value, especially from BaBar.

${\cal S}_{J/\psi K}$ or ${\cal S}_{c\bar cs}$ are traditionally
identified with the CKM phase $\sin2\phi_1/\beta$ as CPV phase in
$B^0$--$\bar B^0$ mixing, since $b\to c\bar cs$ decay should be
dominantly tree level and has very little CPV phase. The issue is
then in comparing direct measurement with the so-called ``CKM"
~\cite{CKM} or ``UT"~\cite{UT} fit results. These studies are able
to fit to all data other than TCPV measurement in $B$ decay, and
``predict" the value for $\sin2\phi_1/\beta = {\cal S}_{J/\psi K}$
($\cong {\cal S}_{c\bar cs}$) assuming 3 generation SM. Before the
Belle 2005 result~\cite{SpsiKBelle}, these predictions tended to
give $\sin2\phi_1/\beta > 0.75$, suggesting a possible ${\cal
S}_{J/\psi K} - \sin2\phi_1/\beta\vert_{\rm fitter}$ problem.

Is there something happening in $b\to c\bar cs$ decay, similar to
what we may be seeing in $\Delta S$ and ${\cal A}_{K^+\pi^0} -
{\cal A}_{K^+\pi^-}$? Unfortunately this problem would be hard to
pin down without adopting an explicit NP model and implementing it
in the fitter. That is, the reference point of
$\sin2\phi_1/\beta\vert_{\rm fitter}$ would itself become the
point of contention and is somewhat ill-defined.

Taking all three hints together, NP phases may well be present in
the $b\to c\bar cs$ amplitude. Surprisingly, the way out may be
DCPV in $B^+\to J/\psi K^+$ mode, thanks to both its precision
measurement nature, and the plausibility of an associated finite
strong phase.


We aim to keep things as simple as possible. Let us put the $B\to
J/\psi K$ decay amplitude in the form
\begin{equation}
{\cal M}(\bar B\to J/\psi \bar K) = a \, e^{i\delta} + b \,
e^{-i\phi} = a \, (e^{i\delta} + \hat b \, e^{-i\phi}),
 \label{eq:ampl}
\end{equation}
where the first term carrying the strong phase is defined with $a$
positive and dominating the rate. We will maintain $a \cong
\sqrt{{\cal B}(B\to J/\psi K)}$ throughout this note, treating the
second term carrying the weak phase as a small correction to the
rate, i.e. $\hat b \equiv b/a \ll 1$. From Eq. (\ref{eq:ampl}), to
good approximation for small $\hat b$, one has
\begin{eqnarray}
 {\cal S}_{J/\psi K^0} &\cong&
  { \sin2\Phi_{B_d} + 2\hat b \sin(2\Phi_{B_d}+\phi)\cos\delta
   \over 1 + 2 \hat b \cos\phi\cos\delta},
 \label{eq:S} \\
 {\cal A}_{J/\psi K^+} &\cong& -2\hat b\sin\phi\sin\delta.
 \label{eq:A}
\end{eqnarray}
The CPV phase in $B_d$ mixing is put in the more general form of
$\Phi_{B_d}$, which would be $\phi_1/\beta$ in SM.
Let us analyze the strength of $\delta$ in general, as well as the
strength $\hat b$ and phase $\phi$ within SM.

The $B\to J/\psi K$ decay is a color-suppressed process, and there
is no factorization theorem. It is generated by the tree level
effective $a_2$ coefficient, hence dominates $a$ in
Eq.~(\ref{eq:ampl}). Since this effective $a_2$ is larger than
naive, it contains hadronic effects that would likely generate a
finite strong phase $\delta \cong \arg a_2$. For color-suppressed
modes that are similarly enhanced, such as $B^0\to D^0\pi^0$,
analysis~\cite{D0pi0} shows that the associated strong phase
$\delta \sim 30^\circ$. Further evidence for such strong phase
comes from experimental study of $B\to J/\psi K^*$ decay. By
angular analysis, strong phase differences between the various
helicity components vary from $\sim 26^\circ$~\cite{dBelle} to
$\sim 34^\circ$~\cite{dBaBar}. Although one is not directly
measuring the strong phase of the longitudinal amplitude (the
analogue of $B\to J/\psi K$), the generic strength is consistent
with the $D^0\pi^0$ result. Thus, we take {\it $\delta \simeq
30^\circ$ as our nominal strong phase for $B\to J/\psi K$ decay}.

Denoting $\Delta_{ct}$ and $\Delta_{ut}$ as the penguin loop
amplitudes associated with the CKM coefficients $\lambda_c \equiv
V_{cs}^*V_{cb}$ and $\lambda_u \equiv V_{us}^*V_{ub}$, in SM
Eq.~(\ref{eq:ampl}) becomes
\begin{eqnarray}
 {\cal M}_{\rm SM} &\simeq&
  a \left[ e^{i\delta}
     + \left\vert{\lambda_u\over \lambda_c}\right\vert
      {\hat\Delta_{ut} \over 1 + \hat\Delta_{ct} }e^{-i\phi_3}
      \right],
 \label{eq:SM}
\end{eqnarray}
where $\phi = \phi_3 = \arg V_{ub}^*$, and $a$ now absorbs
$\Delta_{ct}$, but still saturates the decay rate. Note that to
good approximation $\hat \Delta_{ij}$ is normalized by $a$. We see
that
\begin{eqnarray}
 \hat b_{\rm SM} &\simeq&
  \left\vert{\lambda_u\over \lambda_c}\right\vert
      {\hat\Delta_{ut} \over 1 + \hat\Delta_{ct}},
 \label{eq:bSM}
\end{eqnarray}
is rather small. Since $\lambda_c \Delta_{ct}$ should dominate
$B\to \phi K$ decay~\cite{AphiK}, ignoring kinematic differences,
we have the crude estimate $\vert\hat \Delta_{ct}\vert \sim
\sqrt{{\cal B}(B\to \phi K)/{\cal B}(B\to J/\psi K)} \sim 0.1$.
Taking $\vert{\lambda_u/\lambda_c}\vert \simeq 0.02$ and
saturating $\vert\hat \Delta_{ut}\vert \lesssim \vert\hat
\Delta_{ct}\vert$, we get $\hat b_{\rm SM} = {\cal O}(0.002)$,
which implies a negligible shift in ${\cal S}_{J/\psi K^0}$.
Together with $\phi \simeq \phi_3 \simeq 60^\circ$ and $\vert
\sin\delta \vert \sim 1/2$, we find the order of magnitude
estimate
\begin{equation}
 {\cal A}_{J/\psi K^+}^{\rm SM} = {\cal O}(0.003),
 \label{eq:ASM}
\end{equation}
which can be improved by constraining $\vert\hat \Delta_{ut}/\hat
\Delta_{ct}\vert$.


We now extend to consider New Physics scenarios. For its
simplicity, we keep the form of Eq. (\ref{eq:ampl}) with $a$
saturating the $B\to J/\psi K$ rate, and $b$ the term carrying NP
phase. In actual computations, the SM phase is kept in the
numerics.

It should be clear from Eq.~(\ref{eq:S}) that, assuming
$\sin(2\Phi_{B_d}+\phi^{\rm new})$ and $\cos\delta \sim$ 1, to
have ${\cal S}_{J/\psi K}$ shifted by $\sim$ 0.05 or so, one needs
$\hat b$ to be a couple of percent. From Eq.~(\ref{eq:A}) we see
that ${\cal A}_{J/\psi K}$ could then be a few percent as well.
The question then is not whether a downward shift in ${\cal
S}_{J/\psi K}$ could generate observable ${\cal A}_{J/\psi K}$,
but whether this could be achieved in realistic models.

We will illustrate with New Physics in the electroweak penguin
$P_{\rm EW}$. Since $P_{\rm EW}$ arises from short distance, it
should not have sizable strong phase, so assignment of strong
phase to the $a$ term in Eq.~(\ref{eq:ampl}) is intuitive. We will
consider two realistic models where New Physics CPV phase could
appear in $P_{\rm EW}$: the existence of a fourth generation, or
an extra $Z'$.

In a four sequential generation model, natural for affecting
$P_{\rm EW}$, we have found~\cite{Kpi0HNS}
 1) ${\cal A}_{K^+\pi^0} - {\cal A}_{K^+\pi^-} \neq 0$ can be
    accounted for;
 2) $b\to s\ell^+\ell^-$ rate and $B_s$ mixing can be SM-like;
 3) after satisfying kaon constraints, no visible  deviation of $\Delta m_{B_d}$
   from SM, and $\Phi_{B_d} \sim \phi_1$~\cite{KLpinunu};
 4) $\Delta {\cal S}_{K^0\pi^0,\phi K^0} < 0$ can be generated,
    and is robust against hadronic uncertainties~\cite{HNRS}.
We now illustrate the new point 5): while $\sin2\Phi_{B_d}$ mimics
$\sin2\phi_1/\beta$ (see Fig. 4 of Ref.~\cite{KLpinunu}),
$S_{J/\psi K}$ is in fact driven lower to around 0.7, just at the
right CPV phase value of $\phi_{sb} \simeq 70^\circ$ (defined by
$V_{t's}^*V_{t'b} \equiv r_{sb}e^{i\phi_{sb}}$) for points 1)--4)
above.
We emphasize that point 2) has not changed with the
measurement~\cite{dmsCDF} of $B_s$ mixing; a year ago we stated
that~\cite{Kpi0HNS} $\Delta m_{B_s}$ should be SM-like in 4
generation model.

\begin{figure}[t!]
\smallskip  
\vspace{-3mm}\hspace{-1.5mm}
\includegraphics[width=1.6in,height=0.95in,angle=0]{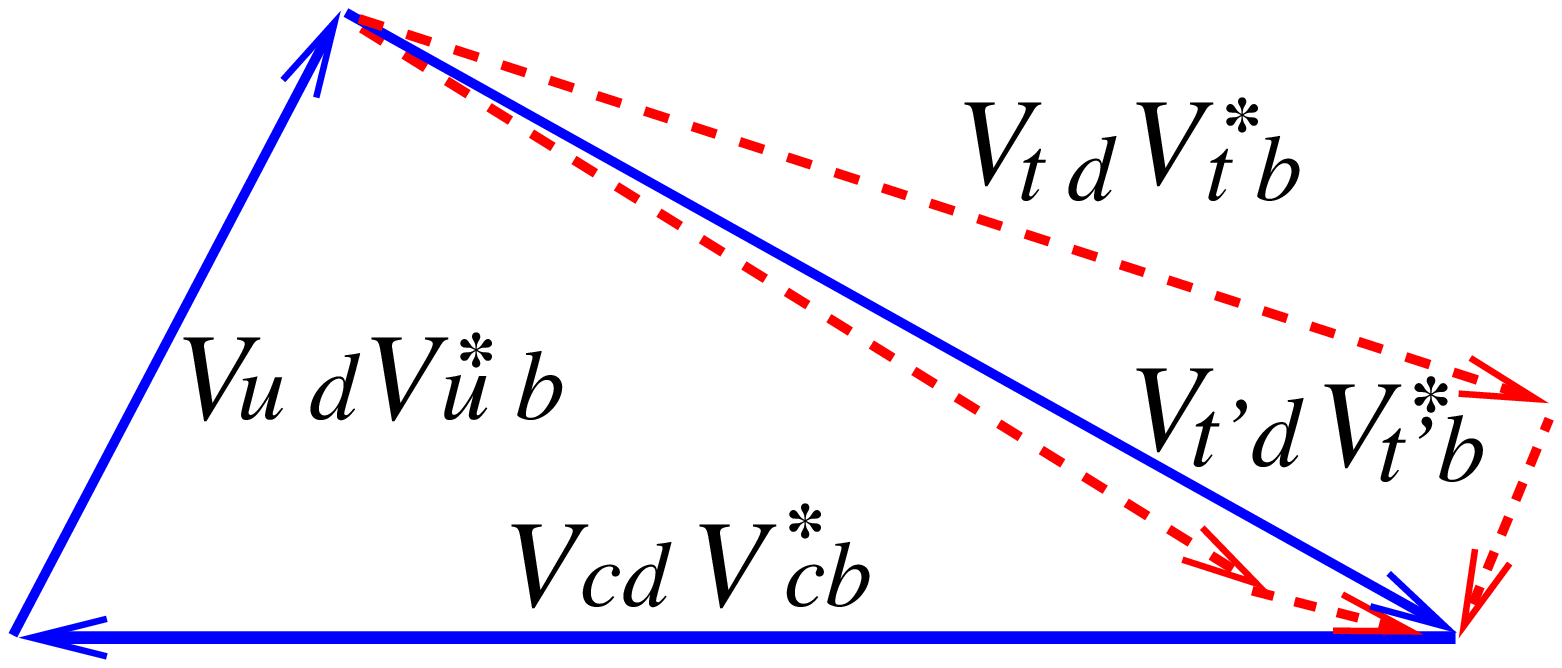}
\hspace{1.5mm}
\includegraphics[width=1.65in,height=1.15in,angle=0]{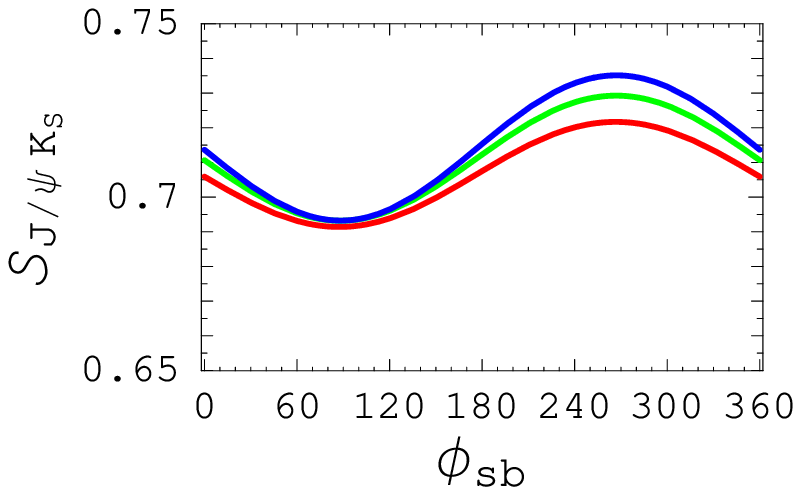}
\vspace{3mm} \hspace{-1.5mm}
\includegraphics[width=1.65in,height=1.15in,angle=0]{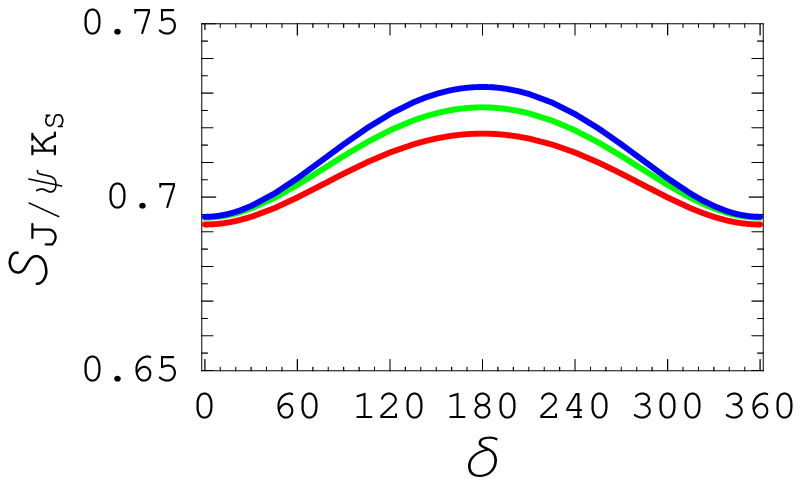}
\vspace{3mm}\hspace{-1mm}
\includegraphics[width=1.65in,height=1.17in,angle=0]{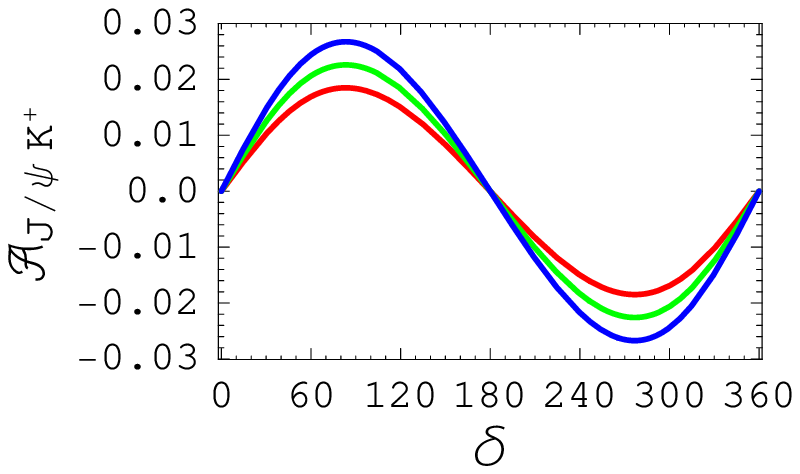}
\vspace{-1mm}
\vskip-0.1cm \caption{(a) Unitarity quadrangle for $b\to d$
transitions with $V_{t^\prime d} V_{t^\prime b}^\ast \equiv r_{db}
\, e^{-i\phi_{db}}$ exaggerated and horizontal scale shrunk by
half, for $\phi_{db} \sim 10^\circ$ and $100^\circ$ (dashed),
compared with 3 generation SM (solid); (b) $S_{J/\psi K^0}$ vs
$\phi_{sb}$, where $V_{t^\prime s}^\ast V_{t^\prime b} \equiv
r_{sb} \, e^{i\phi_{sb}}$; (c) $S_{J/\psi K^0}$ and (d) $A_{J/\psi
K^+}$ vs $\delta$, for $m_{t^\prime}=300$ GeV and $r_{sb}=$ 0.02,
0.025 and 0.03. For (b), we set $\delta=0$. For (c) and (d), we
set $\phi_{sb}=70^\circ$. In (b), (c) and (d), $\phi_{db} =
10^\circ$, and larger $r_{sb}$ gives stronger variation. }
 \label{fig:fourgeneration}
\end{figure}

As discussed, we employ the SM-dominated parameter $a$ to saturate
the $B\to J/\psi K$ rate. The $P_{\rm EW}$ contribution $b$,
arising from 4th generation and carrying CPV phase
$e^{i\phi_{sb}}$, can be computed straightforwardly as for $B\to
K\pi^0$, $\phi K$ modes. One essentially has $b\to s$ transitions
induced by a virtual $Z$, which itself turns into $\pi^0$, $\phi$,
or here the $J/\psi$. We illustrate in
 Fig.~1(a) the $b\to d$ quadrangle.
  Defining $V_{t'd}^*V_{t'b} \equiv r_{db}e^{i\phi_{db}}$, the quadrangle
  with $V_{t'd}V_{t'b}^*$ in 3rd quadrant would lead to a high value of
  $\sin2\Phi_{B_d} \simeq 0.78$, and is disfavored~\cite{KLpinunu}.
  The quadrangle corresponding to $\phi_{db} \sim 10^\circ$ can hardly be
  distinguished from the triangle of the 3 generation case, hence
  mimics SM in $b\to d$ transitions.
 In Fig.~1(b) we plot $S_{J/\psi K}$ vs $\phi_{sb}$
  for our standard parameter set of $m_{t'} \sim 300$ GeV,
  $r_{sb} \sim$ 0.02, 0.025 and 0.03, with strong phase $\delta$ set to zero.
  $S_{J/\psi K}$ can dip below 0.7 around
  $\phi_{sb} \sim 70^\circ$, for all $r_{sb}$.
 Fig.~1(c) illustrates that a small strong phase $\delta$ of order
  $\pm 30^\circ$ or less does not affect this slight suppression of $S_{J/\psi K}$,
  but at 180$^\circ$ would almost restore it back.
 In Fig.~1(d) we plot ${\cal A}_{J/\psi K^+}$ vs $\delta$.
  We see that ${\cal A}_{J/\psi K^+}$ can reach above 1\% for
  $\delta$ up to $\pm 30^\circ$. This could become measurable
  soon.

To illustrate further the correlations between a lower ${\cal
S}_{J/\psi K^0}$ and a finite ${\cal A}_{J/\psi K^+}$, let us
invoke a second model, that of extra $Z'$ boson with flavor
changing couplings. $Z'$ models can also~\cite{Barger} explain the
${\cal A}_{K^+\pi^0} - {\cal A}_{K^+\pi^-}$ difference, as it can
not only enter $P_{\rm EW}$, but affect color-suppressed strong
penguin amplitudes as well. In fact, $Z'$ models are much less
constrained than the previous 4th generation model, since one has
freedom to assign the extra $U(1)$ couplings to various fermions.
For instance, one can ask the $Z'$ to be leptophobic, hence
decouple from $b\to s\ell^+\ell^-$ constraint, while the
consistency of measured $\Delta m_{B_s}$ with SM expectation
allows a range of $Z'$ mass scales~\cite{dmsZ'}. We therefore use
the framework of Eqs.~(\ref{eq:ampl})--(\ref{eq:A}) again to
illustrate first the possible impact on ${\cal S}_{J/\psi K}$.

%
\begin{figure}[t!]
\smallskip  
\vspace{-3mm}\hspace{-1.5mm}
\includegraphics[width=1.64in,height=1.1in,angle=0]{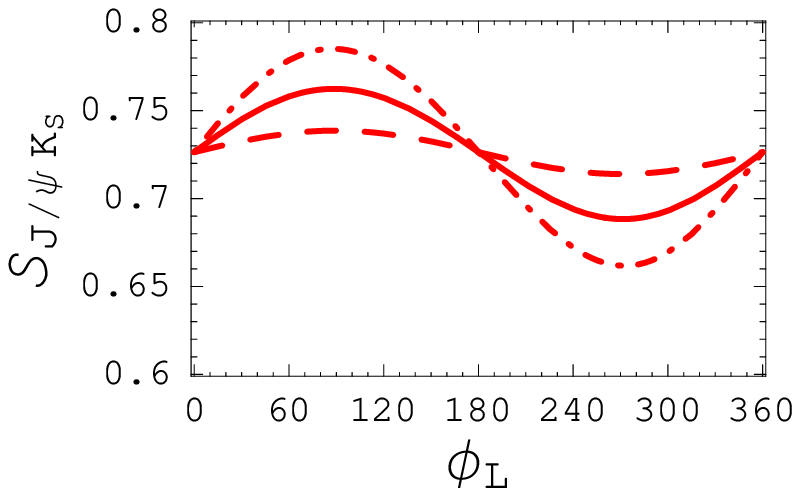}
\vspace{3mm}\hspace{-1mm}
\includegraphics[width=1.64in,height=1.1in,angle=0]{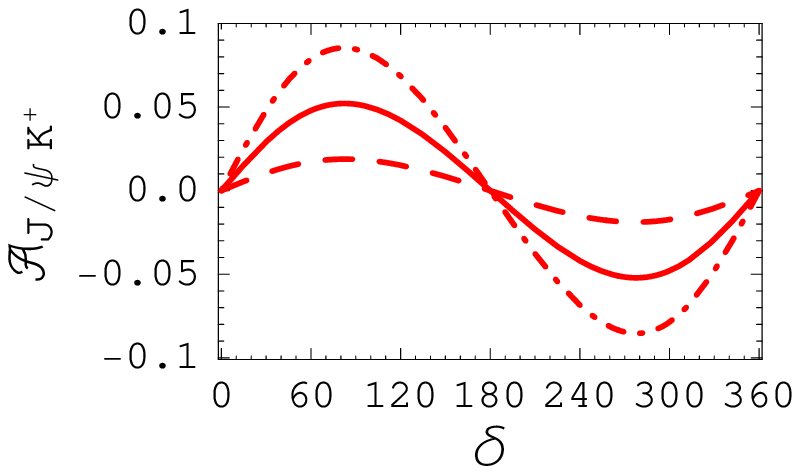}
\vspace{-1mm}
\vskip-0.1cm \caption{(a) $S_{J/\psi K^0}$ vs $\phi_{L}$ for
$\delta=0$, and (b) $A_{J/\psi K^+}$ vs $\delta$ for
$\phi_{L}=-80^\circ$, for $(\xi^{LL}_3,\xi^{LL}_9)= (-0.001,\;
-0.001)$ (dashed), $(-0.003,\; -0.003)$ (solid) and $(-0.005,\;
-0.005)$ (dot-dashed), respectively. }
 \label{fig:Zprime}
\end{figure}


We follow the formalism introduced in Ref.~\cite{Barger}, and
adopt the particular choice for the $Z^{\prime}$ contributions to
the Wilson coefficient at the weak scale. Assuming $LL$ coupling,
the weak phase is placed in the effective charge $\arg B_{sb} =
\phi_L$, and the effective coupling is $\xi_{bsqq} =
c\,B_{sb}B_{qq}$ for $b\to s\bar qq$ transition, where $c =
(g'M_Z/gM_{Z'})^2/|V_{ts}^*V_{tb}|$, and $B_{qq}$ is the effective
$qq$ charge. In $B\to K\pi$ case, for strong penguin one has
$B_{qq} = B_{uu} + 2B_{dd}$, while for EW penguin one has $B_{qq}
= 2(B_{uu} - B_{dd})$. By taking $B_{uu} =
-2B_{dd}$~\cite{Barger}, the contribution would be only in $P_{\rm
EW}$. For our present $B\to J/\psi K$ case, we take $B_{cc} =
B_{uu}$ and there is no $B_{dd}$ contribution, we therefore
parameterize
\begin{eqnarray}
\Delta c_{3}^{\psi K},\ \Delta c_{9}^{\psi K} =
 +\frac{2}{3} \xi^{LL}_3 e^{i \phi_L},\
 +\frac{4}{3} \xi^{LL}_9 e^{i \phi_L},
\end{eqnarray}
where $\Delta c_9^{\psi K}$ contributes to $P_{\rm EW}$, while
$\Delta c_3^{\psi K}$ is a color-suppressed strong penguin, and we
take $\Delta c_5^{\psi K} = \Delta c_7^{\psi K} = 0$.

We plot ${\cal S}_{J/\psi K^0}$ vs $\phi_L$ for $\delta = 0$ in
Fig.~2(a), for three different combinations of the parameters
$(\xi^{LL}_3,\xi^{LL}_9)=(-0.001,-0.001),(-0.003,-0.003)$ and
$(-0.005,-0.005)$.
%
We have checked that a small $\delta$ value does not affect ${\cal
S}_{J/\psi K^0}$, while the parameter choices do not violate
current measurements of $\Delta {\cal S}$. We see that around
$\phi_L \sim -90^\circ$ or so, ${\cal S}_{J/\psi K^0}$ can be
brought down below 0.7, and could even approach 0.65. Keeping
$\phi_L \sim -80^\circ$, we plot ${\cal A}_{J/\psi K^+}$ vs
$\delta$ in Fig.~2(b). We see that for $|\delta| \sim 30^\circ$,
$|{\cal A}_{J/\psi K^+}|$ can be rather sizable, and is
constrained in fact by Eq.~(\ref{eq:expt}). A value of a few \%
could be established with existing data, and by this summer with
over 500M $B\bar B$ events at Belle alone.


We have not intended to go into detailed modelling, but just to
illustrate that the possible hint for New Physics in a lower value
of ${\cal S}_{J/\psi K^0}$, could imply a direct CPV asymmetry
${\cal A}_{J/\psi K^+}$ at 1\% level or higher. Our results imply
that, if one takes the difference of ${\cal S}_{J/\psi K^0} -
\sin2\phi_1/\beta\vert^{\rm fit}$ to be of order $-0.05$, then
${\cal A}_{J/\psi K^+}$ could be as large as a few percent, as
illustrated by the $Z'$ model. This is consistent with the
experimental bound of Eq.~(\ref{eq:expt}). Since the latter is
based on only a fraction of the data that is currently at hand,
there is much room for improvement.

We caution, however, that the strong phase $\delta$, though of
order 20$^\circ$--30$^\circ$ on general grounds, could be small
{\it by accident}. So, ${\cal A}_{J/\psi K^+}$ could still be very
small, even if large NP is present. In addition, for the more
constrained 4th generation model, besides linking all the hints
and giving more wide ranging predictions, the prediction for
${\cal A}_{J/\psi K^+}$ is at the more conservative 1\% level. In
this respect, it is desirable to give a better prediction for
${\cal A}_{J/\psi K^+}^{\rm SM}$ than Eq.~(\ref{eq:ASM}), by
constraining the penguin amplitude ratio $\Delta_{ut}/\Delta_{ct}$
for $B\to J/\psi K$ decay in SM.

Assuming 900M $B\bar B$ events between Belle and BaBar by this
summer, the combined statistical error for ${\cal A}_{J/\psi K^+}$
could reach below 0.005, and a 2\% asymmetry, if present, should
be clearly established. Assuming isospin symmetry,
one could combine with $B^0\to J/\psi K^0$, which requires
tagging. But the error is reduced only by $\sim 10\%$ or so,
unless time dependent analysis provides extra leverage. To probe
$|{\cal A}_{J/\psi K^+}| \sim 1\%$, one needs to await the order
2000M events expected by 2008. With statistical error down to
0.003 or lower, systematic issues would become important. Studies
on systematic errors at the per mill level should be very worthy
towards the era of Super B factory, where many processes of
interest would not be as clean as $B\to J/\psi K$.

We have focused on B factory prospects, but $B\to J/\psi K$ decay
is accessible also at hadronic colliders. With 110 pb$^{-1}$ data
from Tevatron Run I, CDF reconstructed~\cite{psiKCDF} 860 $B\to
J/\psi K^+$ events with $J/\psi \to \mu^+\mu^-$. Scaling to 8
fb$^{-1}$, one expects $6.2 \times 10^{4}$ events towards the end
of Tevatron Run II. If the acceptance and efficiency can be
improved by a factor of 2 or 3, the CDF and D0 experiments could
be competitive with the $\sim 2\times 10^5$ reconstructed $B\to
J/\psi K^\pm$ events per 1000M $B\bar B$ pairs at the B factories.
Systematics could be under better control at hadron colliders due
to more abundant calibration modes.
After 2008, with the advent of the LHC, the ATLAS, CMS and LHCb
experiments should give the best measurement of ${\cal A}_{J/\psi
K^+}$, perhaps even measuring the SM expectation at the few per
mill level.

In summary, in view of possible hints of New Physics in CP
violation in $b\to sq\bar q$ and $c\bar cs$ decays, we point out
that the direct CP violation asymmetry ${\cal A}_{J/\psi K^+}$
could be at the 1\% level or higher, and can be measured in the
next few years. If established, this could become one of the most
convincing evidence for New Physics.

\vskip 0.3cm \noindent{\bf Acknowledgement}.\
 This work is supported in part by NSC 94-2112-M-002-035 and
NSC 94-2811-M-002-053 of Taiwan, and HPRN-CT-2002-00292 of Israel.


\begin{thebibliography}{99}
\bibitem{AphiK}
From ${\cal A}_{\phi K^+} = 0.037 \pm 0.050$~\cite{HFAG}, with
$a\, e^{i\delta} = \lambda_c\Delta_{ct}$ and $b\, e^{-i\phi} =
\lambda_u\Delta_{ut}$, there seems to be little strong phase
difference between $\Delta_{ct}$ and $\Delta_{ut}$.
\end{thebibliography}
\end{document}